\documentclass[manuscript]{aastex}
\usepackage{amsmath}
\usepackage{amsfonts} 
\usepackage[all]{xy}
\usepackage{graphicx}
\usepackage{txfonts}

\shortauthors{Yang et al.}

\begin{document}
\title{Method of Relative Magnetic Helicity Computation II:  Boundary Conditions for the Vector Potentials}

\author{
Shangbin Yang \altaffilmark{1,2}, J\"org B\"uchner \altaffilmark{2},
Jean Carlo Santos \altaffilmark{3},  and  Hongqi Zhang
\altaffilmark{1} }

\altaffiltext{1}{Key Laboratory of Solar Activity, National
Astronomical Observatories, Chinese Academy of Sciences, 100012
Beijing, China}

\altaffiltext{2}{Max-Planck Institute for Solar System Research,
37191 Katlenburg-Lindau, Germany}

\altaffiltext{3} {Laborat\'rio de Plasmas, Instituto de F\'isica,
Universidade de Bras\'lia, Brazil}

\begin{abstract}
We have proposed a method to calculate the relative magnetic
helicity in a finite volume as given the magnetic field in the
former paper (Yang et al. {\it Solar Physics}, {\bf 283}, 369,
2013). This method requires that the magnetic flux to be balanced on
all the side boundaries of the considered volume. In this paper, we
propose a scheme to obtain the vector potentials at the boundaries
to remove the above restriction. We also used a theoretical model
(Low and Lou, {\it Astrophys. J.} {\bf 352}, 343, 1990) to test our
scheme.
\end{abstract}

\keywords{Magnetic helicity}

\section{Introduction}
Magnetic helicity is a key geometrical parameter to describe the
structure and evolution of solar coronal magnetic fields \citep[
e.g.][]{ber99}. Magnetic helicity in a volume $V$ can be determined
as
\begin{equation}
\centering \label{eq:helicityDefine}
 {H_{\rm M}=\int_{V}\mathbf{A}\cdot\mathbf{B}dV},
\end{equation}
where {\bf A} is the vector potential for the magnetic field {\bf B}
in this volume. Magnetic helicity is conserved in an ideal
magneto-plasma \citep{wol58}. As long as the overall magnetic
Reynolds number is large, however, it is still approximately
conserved, even in the course of relatively slow magnetic
reconnection \citep{ber84}. The concept of magnetic helicity has
successfully been applied to characterize solar coronal processes,
for a recent review about modeling and observations of photospheric
magnetic helicity see, {\it e.g.},~\citet{Dem09}. Despite of its
important role in the dynamical evolution of solar plasmas, so far
only a few attempts have been made to estimate the helicity of
coronal magnetic fields based on observations and numerical
simulations (see, {\it e.g.}, Thalmann, Inhester, and Wiegelmann,
2011; Rudenko and Myshyakov, 2011).

\citet{Yang12} developed a method for an efficient calculation of
the relative magnetic helicity in finite 3D volume which already was
applied to a simulated flaring AR \cite{Santos11}. This method
requires the magnetic flux to be balanced on all the side boundaries
of the considered volume. In this paper, a scheme to remove the
restriction has been proposed. In Sec.~\ref{sec:old}, we describe
the restriction of vector potential in the former paper. In
Sec.~\ref{sec:new}, we present the details of the new scheme to
calculate the vector potentials on the six boundaries. In
sec.~\ref{sec:check}, we use the theoretical model to check our
scheme. The  summary and some discussions are given in
Sec.~\ref{sec:summary}.

\section{The former definition of ${\bf A}_{\rm p}$ and {\bf A} at the boundaries}
\label{sec:old}

Let us define a finite three-dimensional (3-D) volume (``box'') in
Cartesian coordinates with a magnetic field ${\bf B}(x, y, z)$ given
in this volume. Let the volume be bounded by $x = [0, l_x]$, $y =
[0, l_y]$, and $z = [0, l_z]$.

First one has to provide the values of ${\bf A}_{\rm p}$ and {\bf A}
on all six boundaries ($x = 0,l_x; y = 0,l_y; z = 0,l_z$). To take
the bottom boundary ($z=0$) for example, we define a new scalar
function $\varphi (x, y)$ that determines  the vector potential
${\bf A}_{\rm p}$ of the potential magnetic field {\bf P} on this
boundary as follows:
\begin{equation}
 {
 A_{\rm p\it x}  =  - \frac{{\partial \varphi }}{{\partial y}}, \qquad
A_{\rm p \it y}  = \frac{{\partial \varphi }}{{\partial x}}, \qquad
 \left.{A_{\rm p \it z} } \right|_{z = 0}  = 0.
 \label{eq:sidebou}
 }
\end{equation}
According to the definition of the vector potential, the scalar
function $\varphi (x, y)$ should satisfy the Poisson equation:
\begin{equation}
\Delta \varphi(x,y)  = B_z (x,y,z = 0).
 \label{eq:bouPoisson}
\end{equation}
The value of $\partial\phi/\partial n$ on the four sides of the
plane $z=0$ is set to zero in Equation~(\ref{eq:bouPoisson}).
According to Eq.(~\ref{eq:sidebou}), $A_{\rm p \it x}$ and $A_{\rm p
\it y}$ will vanish at $y=0, l_y$ and at $x=0, l_x$, respectively,
on the $z=0$ plane. Thus, the corresponding magnetic flux at the
boundary should also vanish because of Amp\`ere's law. The values of
${\bf A}_{\rm p}$ on the other five boundaries could be obtained in
a similar way. For the vector potential {\bf A} at all boundaries
the same values are taken as for ${\bf A}_{\rm p}$. When the
magnetic fluxes at the six boundaries are not zero, we should
calculate the value of vector potentials at the twelve edges of the
three-dimensional (3-D) volume to provide the Neumann  boundary for
the Poisson Equation at each side boundary.
In next section, we will introduce a scheme to calculate the vector
potentials at the twelve edges.


\section{new scheme to obtain $\mathbf A_p$ and $\mathbf A$ at the
boundaries} \label{sec:new}

\begin{figure*}[t]
\centering
 \includegraphics[width=8cm]{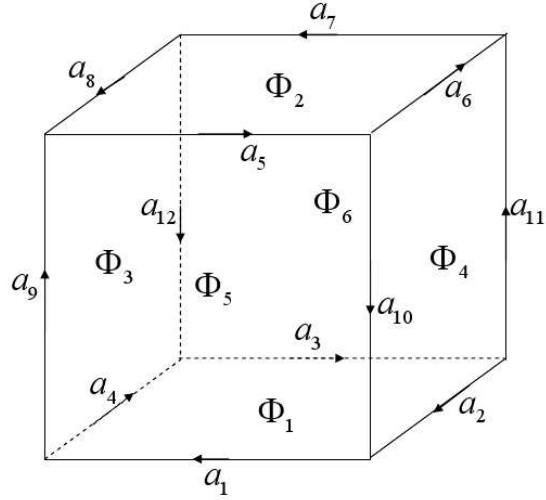}
  \caption{Magnetic flux $\Phi_i~(i=1,...,6)$ at the six boundaries and the integrals $a_i~(i=1,...,12)$ of
$\int{\bf A}_{\rm p} \cdot \rm d{\bf l}$ at the twelve edges.}
  \label{fig:sketch}
\end{figure*}

For the $\mathbf A_p$, we define the magnetic flux
$\Phi_i~(i=1,...,6)$ respectively at each side boundary
($z=0;~z=l_z;~x=0;~x=l_x;~y=0;~y=l_y$). The integrals of $\int{\bf
A}_{\rm p} \cdot \rm d{\bf l}$ at the twelve edges are defined as
$a_i~(i=1,...,12)$. The twelve integrals and the corresponding
directions are represented in Fig.~\ref{fig:sketch}.

According to the Amp\`ere's law, the integral value $a_i$ satisfy
the following linear equations
\begin{equation}
{\rm{TX = B},}
 \label{eq:lineq}
\end{equation}
where $\textrm{B}=(\Phi_1,\Phi_2,\Phi_3,\Phi_4,\Phi_5,\Phi_6)^T$,
$\textrm{X}=(a_1,a_2,a_3,...,a_{12})^T$ and $\textrm{T}$ is a matrix
of 6$\times$12 , which is equal

\begin{equation}
\left[
\begin{array}{cccccccccccc}
1   &  1  & 1  &  1 & 0   & 0   &  0  &  0  &  0  &  0 &  0 &  0 \\
0   &  0  & 0  &  0 & 1   & 1   &  1  &  1  &  0  &  0 &  0 &  0 \\
0   &  0  & 0  &{-1}& 0   & 0   &  0  &{-1} &  1  &  0 &  0 &  1 \\
0   &{-1} & 0  &  0 & 0   &{-1} &  0  &  0  &  0  &  1 &  1 &  0 \\
{-1}&  0  & 0  &  0 &{-1} &  0  &  0  &  0  &{-1} &{-1}&  0 &  0 \\
0   &  0  &{-1}&  0 & 0   &  0  &{-1} &  0  &  0  &  0 &{-1}&{-1}\\
\end{array}
\right]. \label{eq:matrix}
\end{equation}
One can check that the six rows-vector in this matrix is not linear
independent because the magnetic field is divergence free and the
sum of $\Phi_i$ at the six boundaries is zero. Moreover, the unknown
twelve $a_i$ are not unique just under the restriction of above six
conditions. Hence, we need to construct twelve independent
conditions to obtain the unique solution for $a_i$. We define the
new matrix $\hat{\textrm{T}}$ as follows
\begin{equation}
\left[
\begin{array}{ccccccccccccc}
1   &  1  & 1  &  1 & 0   & 0   &  0  &  0  &  0  &  0 &  0 &  0 \\
0   &  0  & 0  &  0 & 1   & 1   &  1  &  1  &  0  &  0 &  0 &  0 \\
0   &  0  & 0  &{-1}& 0   & 0   &  0  &{-1} &  1  &  0 &  0 &  1 \\
0   &{-1} & 0  &  0 & 0   &{-1} &  0  &  0  &  0  &  1 &  1 &  0 \\
{-1}&  0  & 0  &  0 &{-1} &  0  &  0  &  0  &{-1} &{-1}&  0 &  0 \\
1   &  0  &{-1}&  0 &  0  &  0  &  0  &  0  &  0  &  0 &  0 &  0 \\
0   &  1  & 0  &{-1}&  0  &  0  &  0  &  0  &  0  &  0 &  0 &  0 \\
0   &  0  & 1  &  0 & {-1}&  0  &  0  &  0  &  0  &  0 &  0 &  0 \\
0   &  0  & 0  &  1 &  0  & {-1}&  0  &  0  &  0  &  0 &  0 &  0 \\
0   &  0  & 0  &  0 &  1  &  0  & {-1}&  0  &  0  &  0 &  1 &  0 \\
0   &  0  & 0  &  0 &  0  &  1  &  0  & {-1}&  0  &  0 &  0 &  0 \\
0   &  0  & 0  &  0 &  0  &  0  &  1  &  0  & {-1}&  0 &  0 &  0 \\
\end{array} \right]. \label{eq:matrixn}
\end{equation}
One can check that the determinant of $\hat{\textrm{T}}$ is not
zero. According to Cramer rule, the unique solution is existent for
the new linear equation
\begin{equation}
{\rm{{\hat T}X = \hat B},}
 \label{eq:lineqn}
\end{equation}
where $\textrm{X}=(a_1,a_2,a_3,...,a_{12})^T$ and $\hat{
\textrm{B}}=(\Phi_1,\Phi_2,\Phi_3,\Phi_4,\Phi_5,0,0,0,0,0,0,0)^T$.
Then we can obtain the integrals of $\int{\bf A}_{\rm p} \cdot \rm
d{\bf l}$ at the twelve edges. The corresponding vector potential at
the twelve edges could be obtained by using the following equation:

\begin{equation}
\setlength{\jot}{10pt}
\begin{split}
 {\rm{A}}_{{\rm{px}}} \left( {a_i } \right) = \frac{{\pi a_i }}{{2L_x }}\sin ({{\pi x} \mathord{\left/
 {\vphantom {{\pi x} {L_x }}} \right.
 \kern-\nulldelimiterspace} {L_x }}),i = 1,3,5,7 \\
 {\rm{A}}_{{\rm{py}}} \left( {a_i } \right) = \frac{{\pi a_i }}{{2L_y }}\sin ({{\pi y} \mathord{\left/
 {\vphantom {{\pi y} {L_y }}} \right.
 \kern-\nulldelimiterspace} {L_y }}),i = 2,4,6,8 \\
 {\rm{A}}_{{\rm{pz}}} \left( {a_i } \right) = \frac{{\pi a_i }}{{2L_z }}\sin ({{\pi z} \mathord{\left/
 {\vphantom {{\pi z} {L_z }}} \right.
 \kern-\nulldelimiterspace} {L_z }}),i = 9,10,11,12
 \end{split}
 \label{eq:api}
\end{equation}

Note that ${\bf A}_{\rm p}$ at the ends of every edge both are zero
according to the above equation. That is the requirement of
Eq.~(\ref{eq:sidebou}). Then we resolve the Poisson equations to
obtain ${\bf A}_{\rm p}$ at the six boundaries.  For the vector
potential {\bf A} at all boundaries the same values are taken as for
${\bf A}_{\rm p}$. Then we can follow the method of Sec. 2.2 and 2.3
of the former paper \citet{Yang12} to calculate the relative
magnetic helicity in this volume.
\section{Testing the scheme}
\label{sec:check} For testing the new scheme to obtain the vector
potentials at the boundaries, we use the axisymmetric nonlinear
force-free fields of Low and Lou (1990). We used the model labeled
$P_{1,1}$ with $l = 0.3$ and $\Phi = \pi/2$ in the notation of their
paper. We calculated the magnetic field on a uniform grid of
$64\times 64\times 64$. The pixel size in the calculation is assumed
to be 1.

We calculate the magnetic fluxes $\Phi_0$ at the six boundaries and
substitute it to the Eq.~(\ref{eq:lineqn}) to obtain the integral
$a_i$ at the twelve edges of the 3D volume. Then we substitute $a_i$
into Eq.~(\ref{eq:api}) respectively to get the boundary value for
resolving the Poisson equation in Eq.~(\ref{eq:bouPoisson}) at the
six boundaries. After we attain ${\bf A}_{\rm p}$ at the six
boundaries, we could also calculate the magnetic flux $\Phi$
according to the relation between the vector potential and the
magnetic field: ${\bf B}\cdot \hat{\rm n}=\nabla \times {\bf A}_{\rm
p} \cdot \hat{\rm n}$. Table.~\ref{table:check} represents the final
result after we apply the the above scheme. It can be found that the
calculated magnetic fluxes at the six boundaries by using our scheme
respectively coincide well with the original value from the
theoretical model. Note that the total magnetic flux of the
theoretical model is not exact zero. However, it is required that
the total magnetic flux is exact zero when resolving the linear
equation Eq.~(\ref{eq:lineqn}), which cause  the total magnetic
fluxes of $\oint{\bf A}_{\rm p} \cdot \rm d{\bf l}$ and $\Phi$ are
different with that of $\Phi_0$. On the other hand, the numerical
errors when resolving the Poisson equation are also unavoidable,
which will also introduce the difference for the total magnetic flux
as well.

\begin{table*}[T]
\small
\begin{center}
\caption{Testing using the new scheme to a theoretical
model.}\label{table:check}
\begin{tabular}{cccc}
\tableline \tableline
side boundary& $\Phi_0^{\tablenotemark{a}}$ & $\oint{\bf A}_{\rm p} \cdot \rm d{\bf l}^{\tablenotemark{b}}$  & $\Phi^{\tablenotemark{c}}$\\
\tableline
$z=0$   & -3615.81  &  -3615.81    &    -3487.1068 \\
$z=l_z$ &  1461.13  &   1461.13    &     1490.2739 \\
$x=0$   & -1471.95  &  -1471.95    &    -1563.7657 \\
$x=l_x$ & -1471.95  &  -1471.96    &    -1564.9280 \\
$y=0$   &  4006.42  &   4006.42    &     4087.8426 \\
$y=l_y$ & 1068.27   &   1092.17    &    1066.6782  \\
\tableline

Total flux & -23.89   &   0.00012  &    28.99      \\
\tableline
\end{tabular}
 \tablenotetext{a}{Magnetic flux at side
boundary of theoretical model.} \tablenotetext{b}{The integral of
${\bf A}_{\rm p} \cdot \rm d{\bf l}$ for each side boundary.}
\tablenotetext{c}{Magnetic flux from the solution in
Sec.~\ref{eq:sidebou} at each side boundary.}
\end{center}
\end{table*}

\section{Summary}
\label{sec:summary}

In this paper, we propose a new scheme to calculate the vector
potential at the boundaries to remove the restrictions in the former
paper \citet{Yang12}. In principle, now we can calculate the
relative magnetic helicity of any magnetic field structure in
Cartesian coordinates. In the observations, we could use force-free
extrapolation method to obtain the three-dimensional magnetic
structure to analyze the evolution of relative magnetic helicity. On
the other hand, we can also use a sequence of magnetograms to
estimate the accumulated magnetic helicity in the solar corona
\citep[Ref.][]{Dem09}. It will be very interesting to compare the
two types of accumulated magnetic helicity and analyze the
correlation between magnetic helicity and solar eruption
\citep[e.g.][]{Jing12}.
In the simulations, we could also calculate the relative magnetic
helicity directly based the known magnetic field structure to
understand how the magnetic helicity plays an important role in
solar reconnection and dynamos.

\begin{acknowledgements}
This study is supported by grants 10733020, 10921303,
41174153,11173033 11178016 and 11103038 of National Natural Science
Foundation of China, 2011CB811400 of National Basic Research Program
of China, a sandwich-PhD grant of the Max-Planck Society and the
Max-Planck Society Interinstitutional Research Initiative Turbulent
transport and ion heating, reconnection and electron acceleration in
solar and fusion plasmas of Project No. MIF-IF-A-AERO8047.The
authors also like to thank the Supercomputing Center of Chinese
Academy of Sciences (SCCAS) for the allocation of computing time.
\end{acknowledgements}

\end{document}